\documentclass[aps,prl,twocolumn,tightenlines,superscriptaddress,amsmath,amssymb,showpacs,longbibliography]{revtex4-2}

\usepackage{graphicx,graphics,color}
\usepackage[dvipsnames]{xcolor}
\usepackage{amsmath}
\usepackage{amssymb}
\usepackage{amsfonts}
\usepackage{amsfonts}
\usepackage{epstopdf}
\usepackage{bm}
\usepackage{times,xspace}
\usepackage{color}
\usepackage[utf8]{inputenc}
\usepackage{hyperref}
\usepackage{physics}
\usepackage{multirow}
\usepackage{cleveref}
\usepackage[normalem]{ulem}
\usepackage{color,soul}
\usepackage{siunitx}
\usepackage{booktabs}

\DeclareSIUnit\kgf{kgf}
\DeclareSIUnit\kgfpermm{kgf/mm}

\usepackage{verbatim}
\immediate\write18{texcount -nc -inc -sum \jobname.tex > wordcount.tex}

\begin{document}

\title{Edible meta-atoms}

\author{André Souto}
\affiliation{Institute of Physics, University of Amsterdam, 1098 XH Amsterdam, The Netherlands}
\author{Jian Zhang}
\affiliation{Faculty of Mechanical, Maritime and Materials Engineering, Delft University of Technology, Mekelweg 2, 2628 CD Delft, The Netherlands}
\author{Alejandro M. Aragón}
\affiliation{Faculty of Mechanical, Maritime and Materials Engineering, Delft University of Technology, Mekelweg 2, 2628 CD Delft, The Netherlands}
\author{Krassimir Velikov}
\affiliation{Unilever Innovation Centre Wageningen, 6708 WH Wageningen, The Netherlands}
\affiliation{Food Quality and Design, Wageningen University, 6700 AA Wageningen, The Netherlands}
\affiliation{Institute of Physics, University of Amsterdam, 1098 XH Amsterdam, The Netherlands}
\affiliation{Debye Institute for NanoMaterials Science, Utrecht University, 3584 CC, Utrecht, The Netherlands}
\author{Corentin Coulais}
\affiliation{Institute of Physics, University of Amsterdam, 1098 XH Amsterdam, The Netherlands}

\vspace{0.3cm}

\maketitle

\textbf{Metamaterials are artificial structures with unusual and superior properties that come from their carefully designed building blocks---also called meta-atoms. Metamaterials have permeated large swatches of science, including electromagnetics and mechanics. Although metamaterials hold the promise for realizing technological advances, their potential to enhance interactions between humans and materials has remained unexplored. Here, we devise meta-atoms with tailored fracture properties to control mouthfeel sensory experience. Using chocolate as a model material, we first use meta-atoms to control the fracture anisotropy and the number of cracks and demonstrate that these properties are captured in mouthfeel experience. We further use topology optimization to rationally design edible meta-atoms with maximally anisotropic fracture strength. Our work opens avenues for the use of meta-atoms and metamaterials to control fracture and to enhance human-matter interactions.}

Mechanical metamaterials are man-made composites whose architecture provides unique and tunable properties~\cite{Kadic_review,Bertoldi_review,Kochmann_review}. In particular, mechanical metamaterials have shown a wide variety of properties, from enhanced strength-to-weight ratio~\cite{Zheng_NatMat2018} and dissipation~\cite{Shan_AdvMat2015}, to programmable mechanical~\cite{Chen_Nature2021} and shape-changing properties~\cite{Coulais_Nature2016}. A particularly interesting avenue for mechanical  meta-atoms and metamaterials is the design of tunable fracture and strength properties~\cite{Driscoll_PNAS2016,Mitchell_NatMat2016,Zheng_NatMat2018,Bonatti_JPMS2019}. 

Tunable fracture properties have tantalizing prospects for engineering applications where strong and tough structures are much needed. However, little is known about how such tunable fracture could be used to enhance interactions between humans and materials. In particular, little is known about how to use tunable fracture to control mouthfeel and sensory experience upon biting. Controlling sensory experience is an important topic for the design of food products such as soups~\cite{Antoine}, yogurts~\cite{Aguayo_FoodResInt2020}, crackers~\cite{vanEck_BrNutJ2020}, cookies~\cite{Piovesan_Foods2020}, insects \cite{Tan_FoodResInt2017}, as well as emulsions~\cite{Fuhrmann_FoodHydro2020} and protein-based~\cite{Fuhrmann_FoodHydro2020} food products. Although the role of mechanical contrast is generally recognized to influence  mouthfeel~\cite{Santagiuliana_FoodHydro2018}, the use of metamaterials for tunable mouthfeel has remained unexplored. 

\begin{figure}[b!]
\centering
\includegraphics[width=0.9\columnwidth]{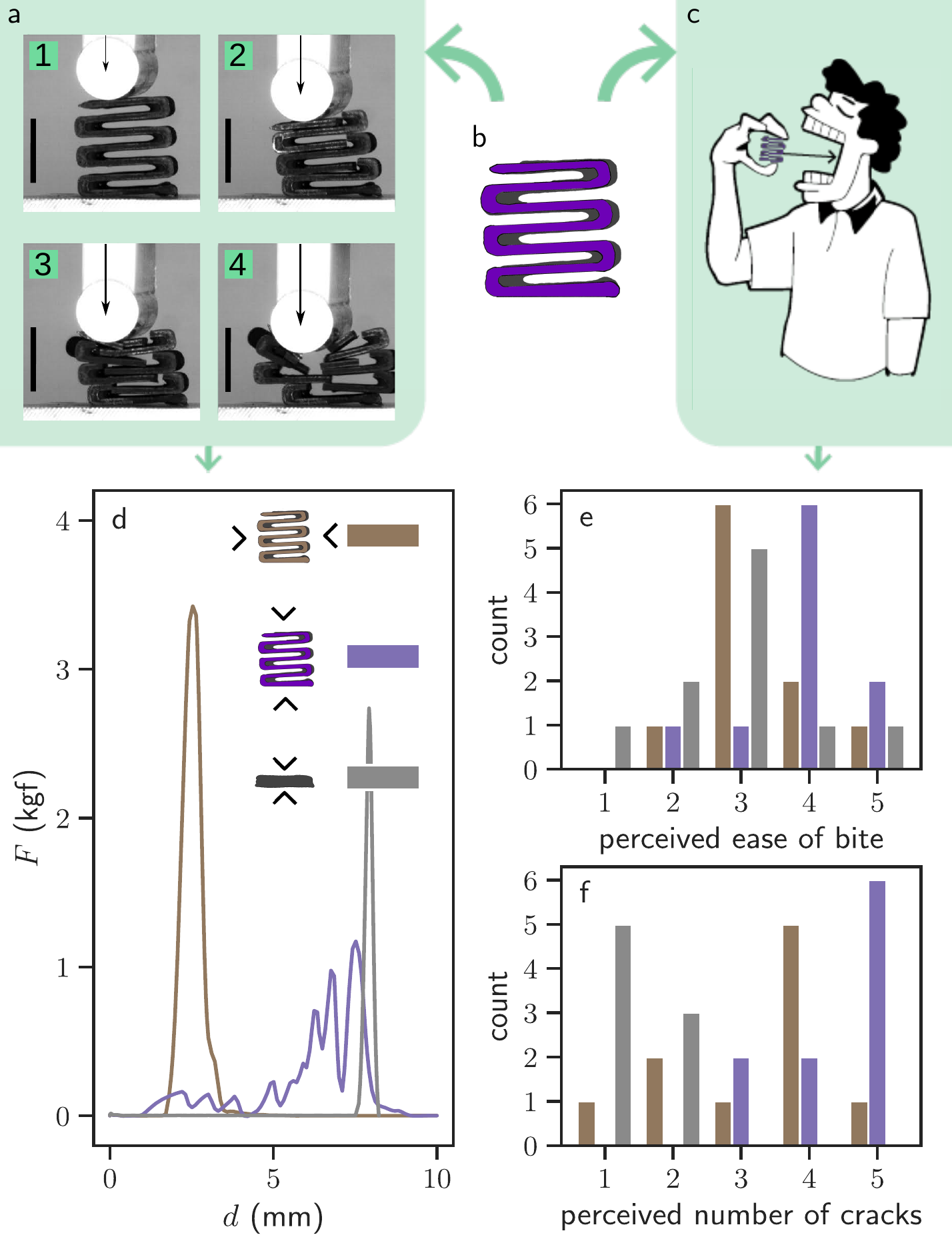}
\caption{
\textbf{Anisotropic edible meta-atoms.}
In panel a, several frames show the step-by-step compression of the $S$-structure shown in b, at a compressive displacement of $\SI{0}{\milli\meter}$ (1), $\SI{3}{\milli\meter}$ (2), $\SI{5}{\milli\meter}$ (3) and $\SI{7}{\milli\meter}$ (4). The scale bar is $\SI{10}{\milli\meter}$. These frames were recorded during a uniaxial compression test, whose results are shown in panel d. These tests were followed up by sensory experiments, as indicated in panel c. Panels e and f show histograms comparing the perceived mechanical properties (ease of bite for panel e and perceived number of cracks for panel f) of the chocolate $S$-structure when bit in two different directions, relative to the reference shape. 
A Fisher test on the data of panel e (resp. f) reveals that there is a $p=3\%$ (resp. $p=9\%$) probability that the ease of bite (resp. the number of cracks) of the sample in the horizontal direction is greater than 3 (resp. 4), while the ease of bite (resp. the number of cracks) of the sample in the vertical direction is smaller than or equal to 3 (resp. 4).}
\label{fig:1}
\end{figure}

\begin{figure*}[t!]
\centering
\includegraphics[width=1.4\columnwidth]{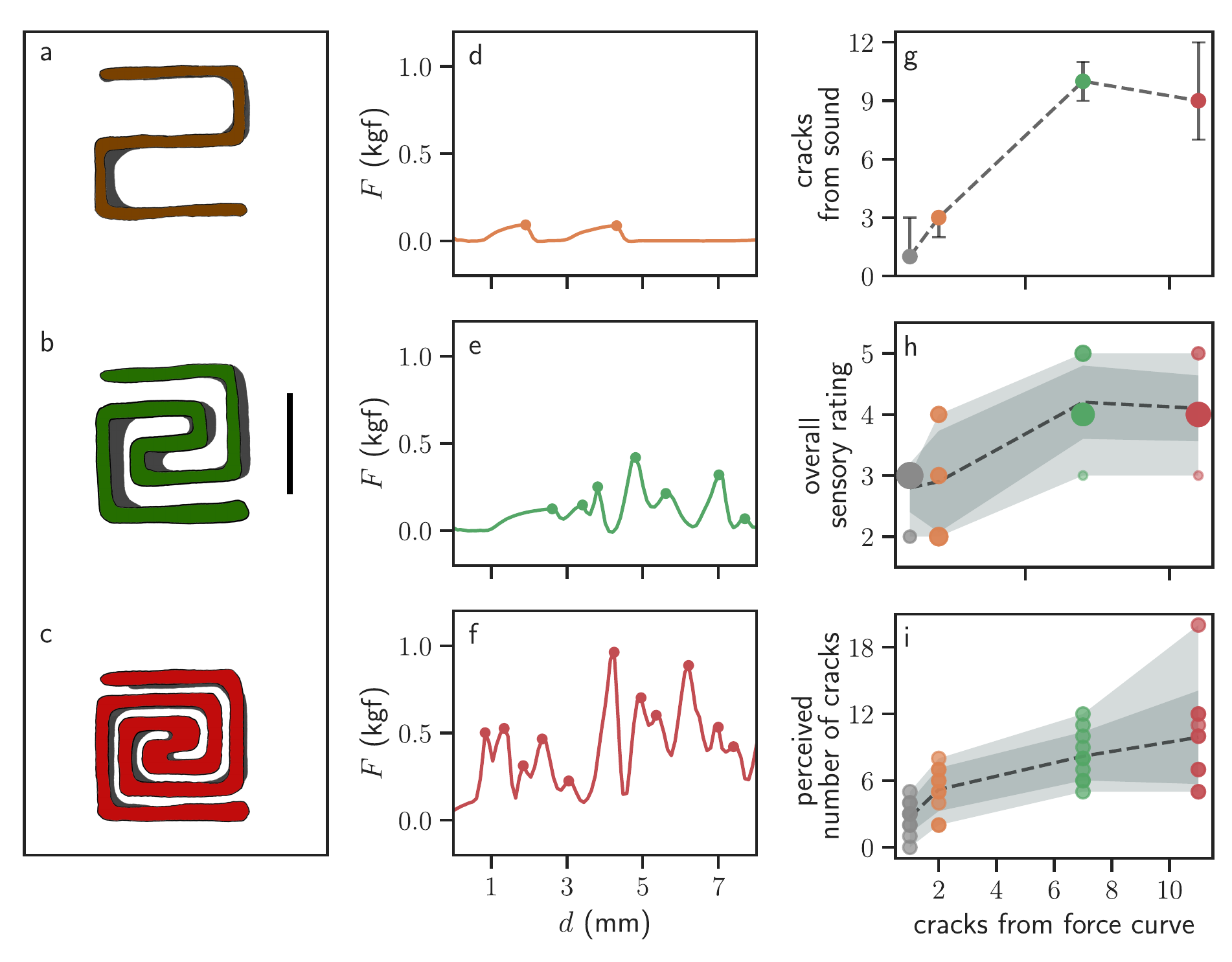}
\caption{\textbf{Edible meta-atoms with tunable numbers of cracks.}
Panels a-c show the spiral-shaped meta-atoms with a winding number $n=1$ (a), $n=2$ (b) and $n=3$ (c). The scale bar is $\SI{10}{\milli\meter}$. Panels d-f show the corresponding force-curves obtained from the uniaxial compression experiments for $n=1$ (d), $n=2$ (e) and $n=3$ (f). Colored dots indicate drops in the applied force, interpreted as the formation of individual cracks. Panels g-i show correlations between the number of cracks extracted from the force-curves and several factors: in g, we plot them against the number of peaks extracted from the audio recordings; in h and i, this data is correlated with the overall rating and the perceived number of cracks, respectively, as assessed by our sensory study. The dashed black line connects the average points and the darker shaded region represents the standard deviation around this average. In panel h, the size of the dot is proportional to the number of occurrences.
Fisher tests on the data of panel i reveal that the probabilities that the meta-atoms with $n=1$ have a perceived number of cracks that is greater than 3 while the perceived number of cracks meta-atoms with $n=2$ and $n=3$ is less than or equal to 3 are both $p=0\%$. The probability that the meta-atoms with $n=2$ have a perceived number of cracks that is greater than 4 while the perceived number of cracks meta-atoms with $n=3$ is less than or equal to 4 is $p=8\%$.}
\label{fig:2}
\end{figure*}

Here we demonstrate that suitably designed edible meta-atoms with controllable fracture properties also have tunable mouthfeel. Namely, we use anisotropic structures to control the ease of bite and use spiral-shaped structures to control the perceived number of cracks. We further demonstrate that topology optimization allows the design of structures with anisotropic fracture properties. Our findings open avenues for design of mouthfeel using edible metamaterials and the topological design of fracture properties.

We start by the $S$-shaped meta-atoms shown in Fig.~\ref{fig:1}b, which we test both mechanically (Fig.~\ref{fig:1}a) as well as with a sensory assessment (Fig.~\ref{fig:1}c). In the present study, we restrict our attention to chocolate as a model brittle material and we use 3D printing to prototype the desired architectures---see \textbf{Materials and Methods} for details on the design approach, the 3D printing, and the mechanical testing protocols. First, the mechanical compression tests reveal that the meta-atom has a very different fracture behavior depending on whether it is compressed along one axis or the other. When the meta-atom is compressed along its horizontal direction, the structure is relatively stiff, strong, of comparable stiffness and strength to that of the reference geometry. It is also brittle, as the force-displacement curve exhibits a single peak. In contrast, when the meta-atom is compressed along its vertical direction, the meta-atom is much softer, as its stiffness (strength) is $40$ ($25$) times lower than in the horizontal direction (see Tab.~\ref{tab:mechanics}). In addition, it is also less brittle, as the force-displacement curve exhibits a finite force for larger range of compressive displacements and multiple peaks. Thus, the meta-atom exhibits a strong anisotropic fracture response, which is difficult to realize in typical food microstructures. 

\begin{figure*}[t!]
\centering
\includegraphics[width=1.7\columnwidth]{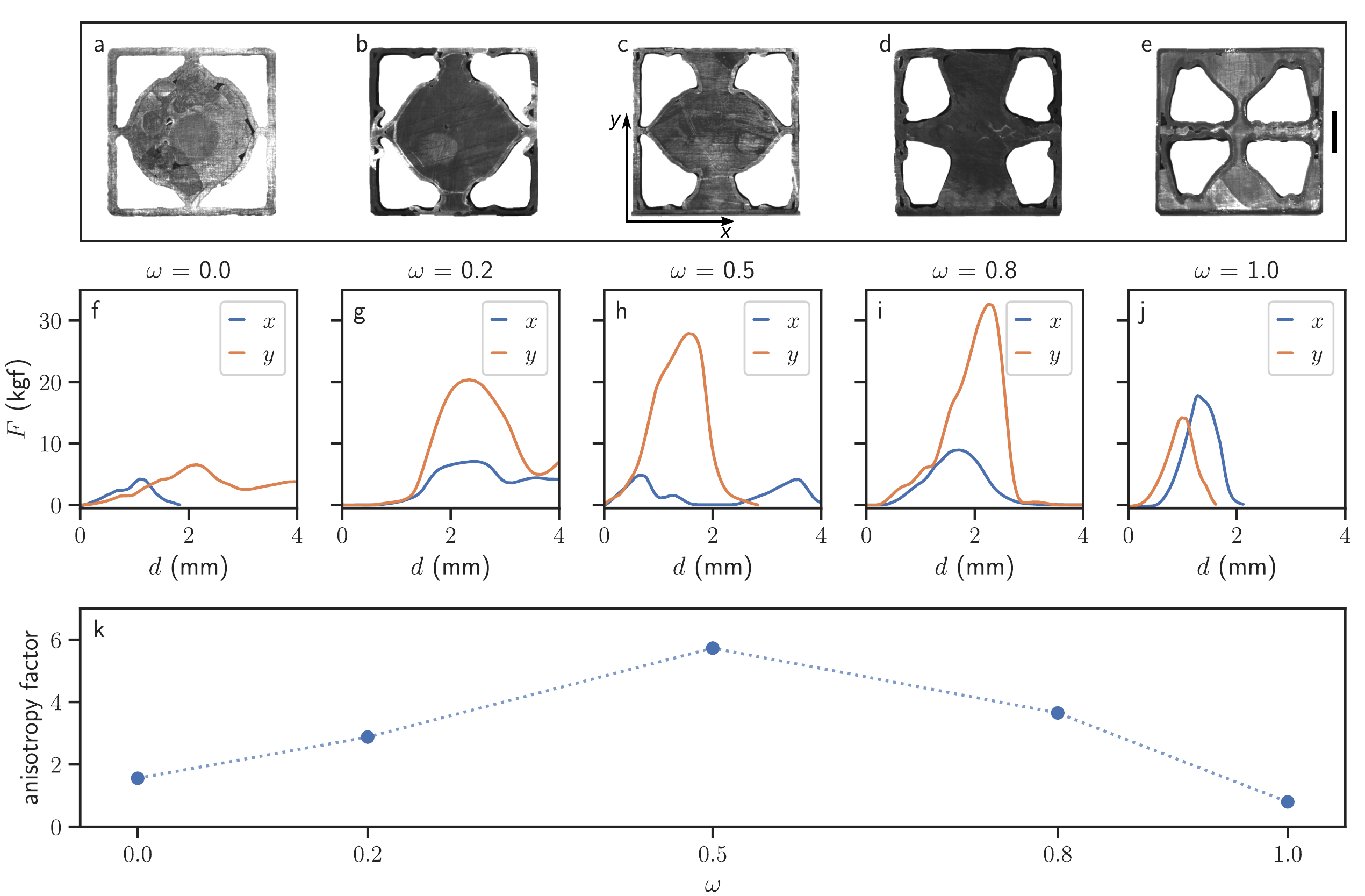}
\caption{\textbf{Design of anisotropic edible meta-atoms with topology optimization.}
Panels (a-e) display the  meta-atoms generated by topology optimization and 3D printed with chocolate. The scale bar is $\SI{10}{\milli\meter}$. Panel (a,b,c,d,e) correspond to the weights $\omega=0,0.2,0.5,0.8$ and $1$, respectively.
We plot  in panels (f-j) the corresponding force-displacement curves extracted from the uniaxial compression tests in the $x$ (blue) and $y$ (orange) directions. Panel (f,g,h,i,j) correspond to the weights $\omega=0,0.2,0.5,0.8$ and $1$, respectively.
Finally, panel g plots the anisotropy factor as a function of the tuning parameter $\omega$. This anisotropy factor is defined as the quotient between the peak forces measured for each direction.}
\label{fig:3}
\end{figure*}

 To investigate whether this stark contrast in mechanical properties between the two orientations leads to a noticeable difference in perceived mouthfeel experience, we perform a sensory assessment on a batch of 10 persons---see \textbf{Materials and Methods} for details about the sensory assessment protocol. The results suggest that a noticeable difference in ease of bite is perceived between the two orientations of the meta-atom (Fig.~\ref{fig:1}e). A Fisher test for statistical relevance reveals that the meta-atom in the horizontal direction is systematically perceived as having an ease of bite lower than 3 whereas the meta-atom in the vertical direction is systematically perceived as having a ease of bite greater than or equal to 3---we find a significance level $p=3\%$. In contrast, a similar Fisher test reveals that both meta-atoms have a comparable perceived number of cracks (Fig.~\ref{fig:1}f, we find a significance level of $p=9\%$, which exceeds the commonly used $p=5\%$ threshold needed to claim statistical relevance~\cite{Wasserman_book}). To conclude, we observe a statistically significant correlation between the anisotropic strength of the meta-atoms and the anisotropy in the ease of bite. An additional interesting feature is that the strong direction of the meta-atom has larger strength and comparable ease of bite than that of the reference, even though the volume fraction of the meta-atom is $64\%$ (see Tab.~\ref{tab:mechanics}), which is significantly lower than the $100\%$ volume fraction of the reference structure.

Which additional mouthfeel experiences can we tune with meta-atoms? Inspired by the multiple local maxima of the meta-atom in the vertical orientation and the slightly larger perceived number of cracks in Fig.~\ref{fig:1}f, we design an additional set of meta-atoms in which the number of fracture and self-contact events can be tuned by geometry. We design a spiral-shaped meta-atom in which the number of windings of the spiral $n$ can be tuned from 1 to 3 (see Fig.~\ref{fig:2}a-c). Upon compression, the force-displacement curve exhibits drastic changes. First, the strength of the meta-atom increases from \SI{8.3e-2}{\kgf} to \SI{4.8e-1}{\kgf}, which is consistent with the fact that the volume fraction increases with the winding number $n$ (see Tab.~\ref{tab:mechanics}). In addition, the force-displacement curves exhibit subsequent local maxima and local minima. Such maxima correspond to local fracture events whereas the local minima correspond to reconfigurations of the meta-atom ensued by the creation of self-contacts. The number of local fractures increases with the winding number $n$.

In addition to quantifying the number of fracture events, we also recorded sound during the experiments and detected fracture events using the sound waveform. Such sound is also expected to be an integral part of the sensory experience during biting~\cite{Zampini_ChemosensoryPercp2010}. We find that the number of fracture events detected with sound exhibits a positive correlation with that of the force-displacement curve (Fig.~\ref{fig:2}g). In addition to the sound recording, we also performed a sensory assessment with these meta-atoms (see \textbf{Materials and Methods} for details on the protocol). We observe that both the overall sensory rating (Fig. \ref{fig:2}h) as well as the perceived number of cracks (Fig. \ref{fig:2}i) exhibit a positive correlation with the number of cracks measured from the force-displacement curve. This positive correlation is partially confirmed by Fisher statistical relevance tests, which confirm the meta-atoms with $n=2$ and $n=3$ have a larger perceived number of cracks than the meta-atom with $n=1$. This is surprising given the large differences in boundary conditions between the mechanical test---realized with ideal boundary conditions---and the sensory assessment, where there is inevitable variability in the loading conditions while biting.

Now that we have shown that meta-atoms with different fracture properties can lead to different mouthfeel experiences and even improve the overall tasting experience, we push their rational design one step further by using topology optimization to create maximally anisotropic structures. We devise an optimization problem that minimizes a quantity $J_1$ when compressing the meta-atom in the vertical $y$ direction and simultaneously maximizes another quantity $J_2$ when compressing it in the horizontal $x$ direction.
$J_1$ and $J_2$ aggregate the energy release rates of every potential crack along the structural boundary. These are then combined into a single objective function $J=\omega J_1 - (1-\omega)J_2$, in which $0 \leq \omega \leq 1$ is a weight parameter. The multiobjective optimization problem has therefore been cast as a single objective problem using a simple weighted sum---see \textbf{Materials and Methods} for details about the computational approach.

We run our topology optimization algorithm for multiple values of $\omega$ with a fixed volume fraction of $50\%$. For a value $\omega = 0$, only the second term in $J$ is non-zero, resulting in an optimized topology that is very brittle when compressed along $x$. Conversely, for $\omega = 1$ only the first term in $J$ is non-zero, with an optimized topology that makes the meta-atom very tough along $y$. Although the topology optimization algorithm optimizes for fracture resistance, since chocolate is brittle, we expect the enhanced fracture anisotropy to also manifest itself via an enhanced strength anisotropy~\cite{Munro_JAmCerSoc1999}. Five optimized designs were 3D-printed in chocolate (Fig.~\ref{fig:3}a) and their mechanical properties were tested (Fig.~\ref{fig:3}b-f)---See \textbf{Materials and Methods} for printing and test protocols.
The mechanical tests reveal that all structures have a similar force-displacement curve. Initial stiffness and maximum strength values are reported in  Tab.~\ref{tab:mechanics}; the maximum strength was obtained at compressive displacements between $1$ and $\SI{2}{\milli\metre}$. 
As expected, we find that the meta-atom displays a maximally anisotropic strength for $\omega=0.5$ (Fig.~\ref{fig:3}).  This result can be interpreted from the geometry of the meta-atom, which has thin necks in the $x$ direction and thick walls in the $y$ direction. 
Such strong anisotropy is twice as large as that of the structure in Fig.~\ref{fig:1}, therefore we expect it to have to a strong ease of bite anisotropy.  

To conclude, by designing, performing mechanical tests and carrying out sensory assessments of edible meta-atoms, we have shown that it is possible to enhance the mouthfeel sensory experience. Our work opens the door to rational design of human-matter interactions and of fracture properties by using optimization methods for fracture, for edible products, but also engineering structures and beyond.

\bibliographystyle{naturemag}

\emph{Acknowledgments.---} We thank A. Deblais for insightful discussions and suggestions, S. Koot and D. Giesen for technical assistance. We acknowledge funding from an IXA Physics2market grant. C.C. acknowledges funding from the European Research Council via the Grant ERC-StG-Coulais-852587-Extr3Me. All the codes and data supporting this study are available on the public repository https://doi.org/10.5281/zenodo.4632399.

\clearpage
\setcounter{equation}{0}
\renewcommand{\theequation}{S\arabic{equation}}%
\setcounter{figure}{0}
\renewcommand{\thefigure}{S\arabic{figure}}%
\setcounter{table}{0}
\renewcommand{\thetable}{S\arabic{table}}%

\section{Materials and Methods}

\subsection{Design approach}

The designs found on \cref{fig:2}a were created in different fashions. The various spirals were drawn in FreeCAD, an open-source and freely available CAD software package, whereas the $S$-structure is simply one of the predefined patterns available in Visual Machines, the control software for EnvisionTEC's 3D-Bioplotter. Whereas the latter contains several parallel beams, giving rise to its anisotropic properties, the spirals' interesting mouthfeel is less dependent on the biting direction. Overall, both designs rely on the same principle: the addition of breakable elements such as beams perpendicular to the direction of compression should result in a richer feeling in the mouth.

Not only do the spirals possess a large number of breakable elements, the variable winding number ($n$) provides an obvious tuning parameter to find the sweet spot in terms of sensory perception. Initially, we tested winding numbers up to $n=7$ for larger pieces but when printing bite-sized samples for the sensory study, we realized that $n>3$ wasn't printable given the limited resolution---see \textbf{3D printing} for an explanation. In the end, we arrived at $15\times15\times10$ \si{\milli\meter\cubed} for the dimensions of the spirals and the $S$-structure. The reference shape is $15\times15\times4$ \si{\milli\meter\cubed}, making it thinner than the other shapes. This was done in an attempt to match the overall amount of chocolate deposited between different geometries.

\subsection{3D printing}
3D printing chocolate proved to be a challenging endeavor, due to the material's polymorphism~\cite{hao2010}. There are 6 different crystal forms of chocolate, each with their own mechanical properties and melting points. Typical chocolate products favor form V crystals, which tend to be shiny and smooth, producing audible snaps when bitten into~\cite{Beckett2008}. This particular crystal form melts at \SI{34}{\celsius}~\cite{SCHENK2004} so the working temperature of the chocolate must be kept under this value. We used Cacao Barry's Venezuela 72\% dark chocolate, due to its advertised fluidity and high concentration of cacao---lower purity levels would further complicate the tempering and printing processes.

Tempering was achieved using the seeding method, which consists of the following steps:

\begin{itemize}
    \item Bring the chocolate temperature above the highest melting point of its different crystal forms (\SI{36}{\celsius} for form VI crystals). To ensure fully melted chocolate, we heated the material up to \SI{45}{\celsius}. 
    \item When the desired temperature is reached, turn the heating device off.
    \item Proceed to gradually add solid pellets of pre-tempered chocolate, stirring carefully, while monitoring the temperature of the liquid.
    \item When the temperature drops below the aforementioned melting point for form V crystals (\SI{34}{\celsius}), the chocolate is now ready to be printed.
\end{itemize}

The seeding process ensures that the right crystals are dispersed into the mix, acting as nucleation sites for further growth of the desired crystal form.

After tempering, the chocolate is loaded onto a syringe which is then inserted in the 3D-Bioplotter's heated cartridges, which are kept at \SI{32}{\celsius}. To make sure that, after deposition, solidification happens as quickly as possible, the printing base is kept at \SI{12}{\celsius} and a simple fan is used to create air flow. 

In extrusion-based 3D printing, smaller nozzles are preferred due to the higher resolution they allow for, but we quickly realized that if the nozzle was too small, it would easily get blocked by the growing nucleation sites dispersed in the chocolate. So, we had a relatively large (\SI{1.2}{\milli \meter}) nozzle custom made, which worked effectively most of the time---we circumvented the limited resolution by working with single walled designs.

The 3D-Bioplotter uses air pressure to extrude the printing material. This means that a balance between air pressure and nozzle speed must be struck. We found that there was no consistent way to strike this balance, as the fluidity of chocolate seems to fluctuate when it is left unstirred inside the syringe. A printing session would often start with relatively low pressures (\SI{0.1}{\bar}) and high speeds (\SI{40}{\milli \meter \per \second}) but, as the chocolate thickened, we were forced to increase the pressure and to reduce the speed. Needless to say, the printing process requires frequent calibration to ensure that the printed lines have approximately constant thickness---matching the diameter of the nozzle.

\subsection{Mechanical testing}

Testing was done using an Instron 5943 uniaxial compression device. The samples were placed on a flat bed and crushed by a cylindrical rod attached to the movable arm of the machine; a constant compression speed of \SI{5}{\milli \meter \per \second} was set. A \SI{500}{\newton} load cell was used to register the force applied by the aforementioned cylindrical rod. 

During the compression, a Basler acA2040-90um camera was used to record pictures of the process, while a MiniDSP UMIK-1 microphone was used to capture audio. 

The different samples were kept refrigerated right until the moment of testing, to ensure temperature uniformity across all tests. 

\begin{table}[t!]
\begin{tabular}{l c c c l}
\toprule
  & \textbf{Strength} & \textbf{Stiffness } & \textbf{Volume} &  \\ 
 \textbf{Sample} & (\si{\kgf}) & (\si{\kgfpermm}) & \textbf{ Fraction} &  \\ \midrule
\textbf{S-structure ($x$)}            & 3.41                    & 4.11                        & 0.64                     &  \\ 
\textbf{S-structure ($y$)}            & 0.148                   & 0.112                       & 0.64                     &  \\ 
\textbf{Reference}                    & 2.66                    & 6.19                        & 1.0                      &  \\ 
\textbf{Spiral ($n=1$)}               & 0.0832                  & 0.0761                      & 0.32                     &  \\ 
\textbf{Spiral ($n=2$)}               & 0.108                   & 0.0613                      & 0.48                     &  \\ 
\textbf{Spiral ($n=3$)}               & 0.484                   & 0.364                       & 0.64                     &  \\ 
\textbf{TO, $\omega = 0.0$ ($x$)} & 4.23                    & 3.99                        & 0.50                     &  \\ 
\textbf{TO, $\omega = 0.0$ ($y$)} & 6.47                    & 3.04                        & 0.50                     &  \\ 
\textbf{TO, $\omega = 0.2$ ($x$)} & 7.00                    & 3.89                        & 0.50                     &  \\ 
\textbf{TO, $\omega = 0.2$ ($y$)} & 20.3                    & 11.6                        & 0.50                     &  \\ 
\textbf{TO, $\omega = 0.5$ ($x$)} & 4.76                    & 7.41                        & 0.50                     &  \\ 
\textbf{TO, $\omega = 0.5$ ($y$)} & 27.7                    & 17.6                        & 0.50                     &  \\ 
\textbf{TO, $\omega = 0.8$ ($x$)} & 8.84                    & 6.08                        & 0.50                     &  \\ 
\textbf{TO, $\omega = 0.8$ ($y$)} & 32.6                    & 16.3                        & 0.50                     &  \\ 
\textbf{TO, $\omega = 1.0$ ($x$)} & 17.7                    & 21.4                        & 0.50                     &  \\ 
\textbf{TO, $\omega = 1.0$ ($y$)} & 14.2                    & 15.4                        & 0.50                     &  \\ \bottomrule
\end{tabular}
\caption{Summary of some mechanical properties of the different designs.}
\label{tab:mechanics}
\end{table}

\subsection{Sensory assessment}

\begin{table*}[t!]
	\begin{tabular}{l cc cc cc cc cc cc cc }
		\toprule
		\multirow{2}{*}{\textbf{Sample}} & \multicolumn{2}{|c|}{\textbf{Reference}} & \multicolumn{2}{c|}{\textbf{Sp. ($n=1$)}} & \multicolumn{2}{c|}{\textbf{Sp. ($n=2$)}} & \multicolumn{2}{c|}{\textbf{Sp. ($n=3$)}} & \multicolumn{2}{c|}{\textbf{S-str. ($x$)}} & \multicolumn{2}{c|}{\textbf{S-str. ($y$)}} & \multicolumn{2}{c|}{\textbf{S-str. ($z$)}}                             \\\cline{2-15} 
                            & \multicolumn{1}{|c}{Av.} & \multicolumn{1}{c|}{Dev.} & Av.& \multicolumn{1}{c|}{Dev.} & Av.& \multicolumn{1}{c|}{Dev.} & Av.& \multicolumn{1}{c|}{Dev.}& Av.& \multicolumn{1}{c|}{Dev.}& Av.& \multicolumn{1}{c|}{Dev.}& Av.& \multicolumn{1}{c|}{Dev.}                   \\ \midrule 
\textbf{Crunchiness}        & 2.3 & 0.6 & 3.1 & 1.2 & 4.0 & 0.9 & 4.2 & 0.9 & 3.7 & 0.9 & 4.0 & 1.0 & 3.3 & 1.0             \\
\textbf{Ease of bite}       & 2.9 & 1.0 & 4.3 & 0.9 & 4.1 & 0.5 & 3.5 & 0.8 & 3.3 & 0.8 & 3.9 & 0.8 & 3.1 & 1.1             \\
\textbf{Normalized cracks}  & 1.1 & 0.6	& 2.2 &	0.6 & 3.6 & 0.8	& 4.1 & 0.9	& 3.2 & 1.1	& 4.3 & 1.0	& 2.4 & 0.9             \\
\textbf{Overall rating}     & 2.8 & 0.4 & 2.9 & 0.8 & 4.2 & 0.6 & 4.1 & 0.5 & 3.5 & 0.5 & 4.0 & 0.4 & 3.0 & 1.0             \\ \bottomrule
	\end{tabular}
\caption{Results obtained from the sensory study, summarizing how the different shapes were perceived, across different categories: crunchiness, ease of bite and overall rating, on a scale of 1-5, and the absolute estimated number of cracks. The estimated number of cracks was then normalized according to \cref{eq:norm}. We present the averaged values across all participants in the "Av." subcolumns and the standard deviations in the "Dev." subcolumns.}
\label{fig:table}
\end{table*}

The sensory study was performed by a panel of 10 untrained volunteers, recruited via email. The group was composed of students, researchers and technical staff, all unaware of the purpose of the study. Upon arrival in the testing room, they were provided with a printed questionnaire, containing instructions on how to handle and bite the pieces using their molars. The participants were also instructed to close their jaws at a deliberately slow pace, in an effort to match the conditions of the sensory study to those of the mechanical tests. Furthermore, the different samples were kept refrigerated until the moment of testing, spending only a few seconds exposed to the room temperature before being ingested. The order in which the samples were presented was randomized and each was assigned a codename, so as not to introduce any biases in its perceived properties. Between trials, the participants were allowed to take breaks to cleanse their palates with tap water. The participants were allowed to try each piece more than once, if necessary, to get a better sense of its characteristics. The participants were asked to focus specifically on the first bite, filling in their answers for each question based on their impressions of this experience~\cite{D0FO01787F, Antoine}. They were asked to rate each sample with a number between 1 and 5 for each of the following questions:

\begin{itemize}
    \item How crunchy was it?
    \item How easy was it to bite?
    \item How would you rate the overall experience?
\end{itemize}

Finally, the participants were asked to estimate the absolute number of cracks felt. The results for this particular question were then normalized as follows:

\begin{equation}
    \bar{a}_i = 5\times\frac{a_i}{\max{(a_1, a_2,...)}}
    \label{eq:norm}
\end{equation}

Here, $\bar{a}_i$ and $a_i$ are, respectively, the normalized and absolute values of the estimated number of cracks for a given sample $i$. This normalization was chosen to take into account different individual criteria for what counts as a crack.

\subsection{Topology optimization}

\begin{figure*}[!htb]
             \centering%
 	\def\svgwidth{0.75\textwidth}
        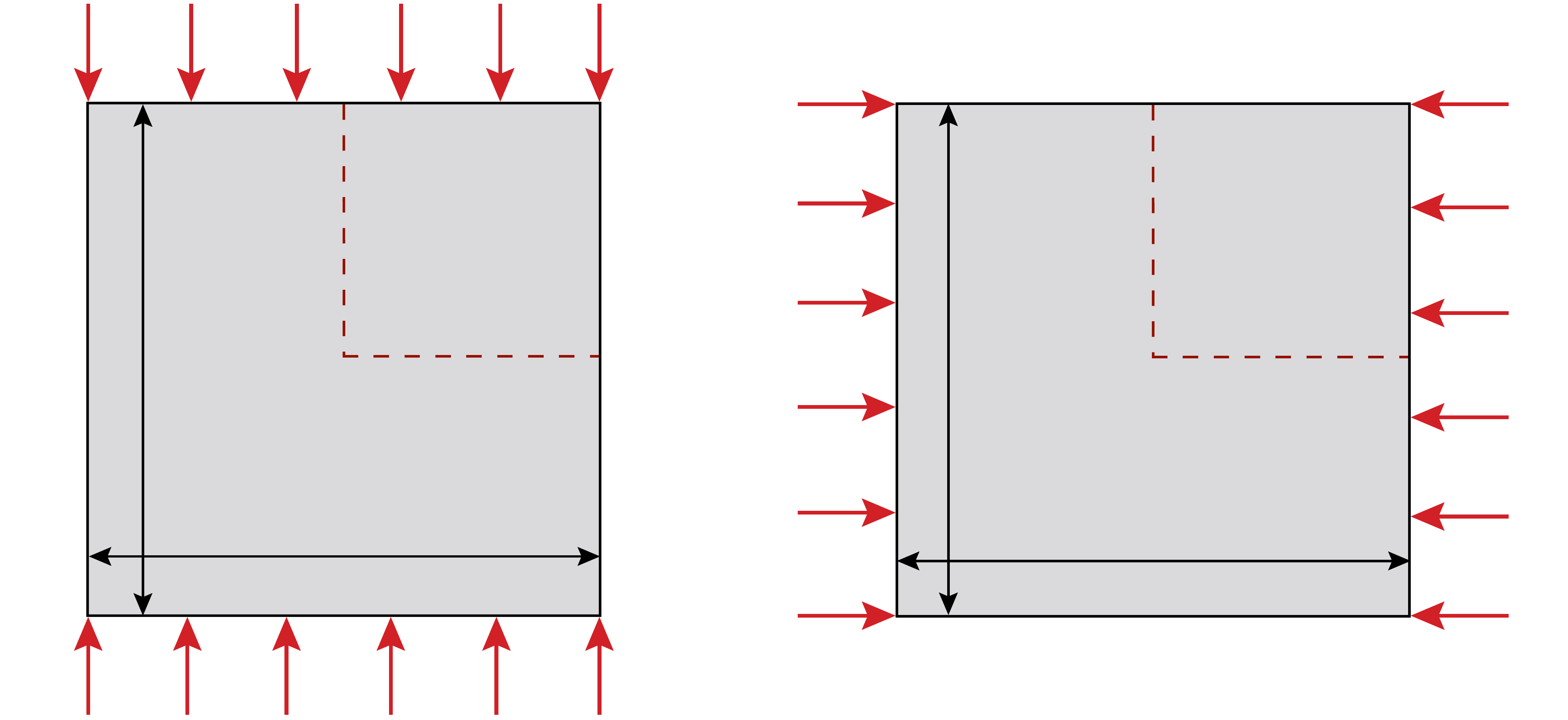	
\caption{A domain with dimension $2 \times 2$ under compression, where $\boldsymbol t_1$ and $\boldsymbol t_2$ are prescribed on the vertical and horizontal directions, respectively. Under the finite element analysis, a quarter of domain (marked with red dashed segments) is considered.} 
\label{fig:loading_schematic} 
\end{figure*}

\begin{figure*}[!htb]
\centering
\def\svgwidth{0.75\textwidth}
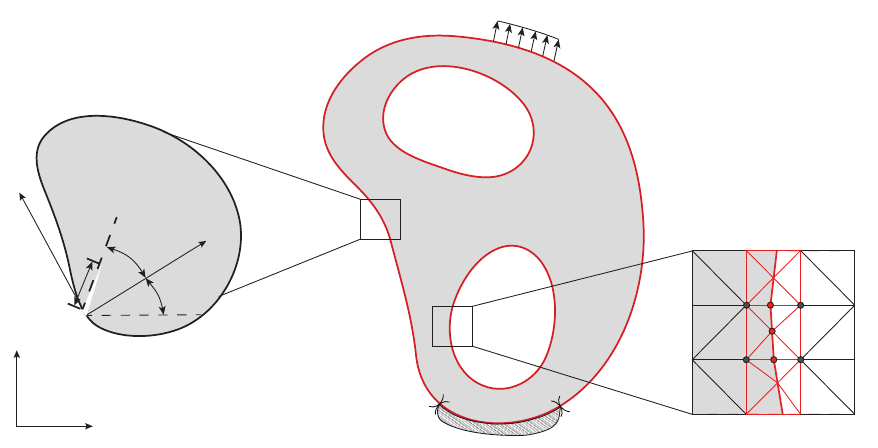
\caption{The illustration of a crack with length $\eta$ nucleating at node $j$, where $\gamma$ is the angle between this crack and the internal normal of structural boundary, and $\beta$ is the angle between the global coordinate system and local coordinate system located at node $j$. Enriched nodes (marked with red circles) are detected at intersections with between structural boundaries and the edge of elements, and integration elements are created near the boundary.} 
\label{fig:angle}
\end{figure*}

Topology optimization (TO)~\cite{Bendsoe:1988,Sigmund:1994} is a technique widely used in structural optimization that combines finite element analysis (FEA) with a gradient-based optimizer for obtaining optimal designs. Starting with a computational domain, the procedure seeks the most appropriate material layout to minimize a given objective function subject to given constraints. In this work we use TO to introduce anisotropy into the meta-atom structure, for which we use two loading cases as shown in Fig.~\ref{fig:loading_schematic}. Due to symmetry, we consider only a quarter of the domain during the optimization. Our objective function thus minimizes some measure of the energy release rate $J_1$ when compressing the meta-atom in the vertical direction, while maximizes another value $J_2$ when compressing along the horizontal direction. The optimization, which is also subject to a target $V_c = 50\%$ volume constraint of chocolate, is mathematically defined as
\begin{equation}
\begin{split}
\text{minimize }  & J  = \omega J_1 - (1-\omega) J_2 \\
\text{such that } & \quad \boldsymbol K_1 \boldsymbol U_1 = \boldsymbol F_1, \\
& \quad \boldsymbol K_2 \boldsymbol U_2 = \boldsymbol F_2, \\
 & \quad V_{\mathrm{solid}} \leq V_c,
\end{split}
\end{equation}
where $J_i$ is computed after solving its associated discrete finite element equilibrium equation $\boldsymbol {K}_i \boldsymbol{U}_i = \boldsymbol{F}_i, i=1, 2$, $0 \leq \omega \leq 1$ is a weight factor that essentially transforms a two-objective optimization problem into a single one, and $V_{\mathrm{solid}}$ is the volume fraction of solid material.
After solving the $i$th discrete problem, $J_i$ is computed as
\begin{align}
J_i = \frac{1}{N}\sum_{j=1}^{N} G_{j},
\end{align}
where $N$ is the number of nodes along chocolate-void interfaces, and $G_{j}$ is the energy release rates of the $j$th node, which is approximated using topological derivatives~\cite{Silva:2011} as
	\begin{equation} \label{eq:energy_release_rate}
G_{j} = \frac{\pi\eta}{\overline{E}} \boldsymbol{\sigma}_j^{\intercal} \boldsymbol Q(\gamma, \beta) \boldsymbol{\sigma}_j.
	\end{equation} 
In this equation, $\boldsymbol{\sigma}_j$ is Cauchy's stress evaluated at a node, and $\overline{E}=E /\left(1-\nu^{2}\right)$ for plane strain conditions, in which $E$ is Young's modulus and $\nu$ is Poisson’s ratio. As shown in Fig.~\ref{fig:angle}, a crack nucleating at a chocolate-void interface has length $\eta$ and angle $\gamma$, the latter measured from the normal vector to the interface; $\beta$ is the angle between the global coordinate system (marked in red) and the local coordinate system positioned at node $\boldsymbol{x}_j$. Finally, $\boldsymbol{Q}$ is a matrix that is function of angles $\gamma$ and $\beta$, relating the two coordinate whose details are given the appendix.

Our approach to topology optimization~\cite{Boom:2021aa} uses a level set function $\phi$, whose zeroth value represents the interface between chocolate and void, an enriched finite element formulation to accurately determine the structural response, and the method of moving asymptotes (MMA) as the optimizer to update the design variables. Noteworthy, energy release rates are computed at the locations of enriched nodes that are added to the finite element formulation to properly describe the discontinuous kinematics at void-solid interfaces. In comparison with standard density-based TO methods, our approach can provide smooth and crisp designs without the need for post-processing~\cite{Boom:2021aa}.

\begin{figure*}[!htb]
\centering
 	\def\svgwidth{0.75\textwidth}
        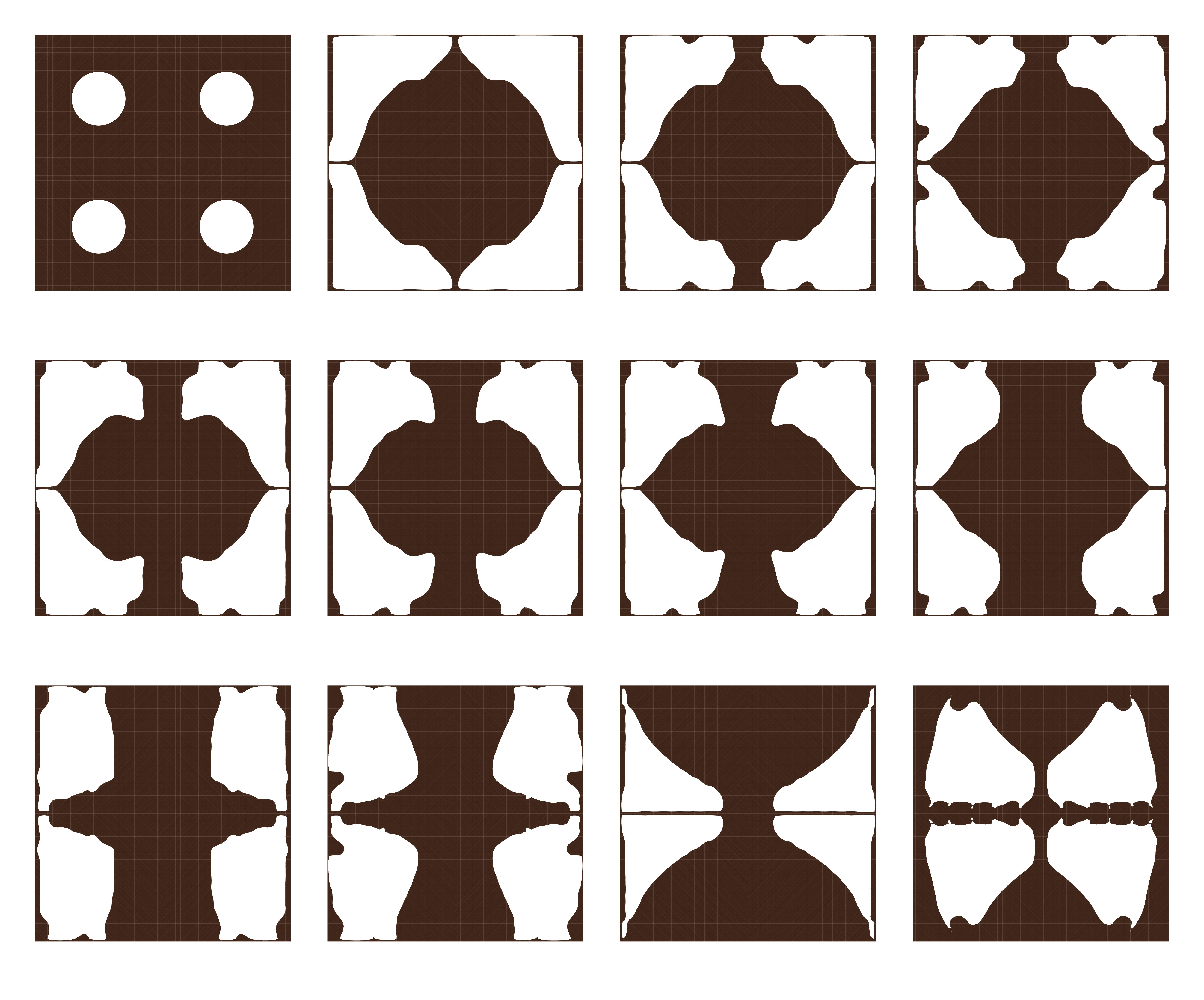	
\caption{(a) Initial design with four holes in the whole design domain; (b) - (l) Final designs with different weights $\omega = 0, 0.1, ..., 1$} 
 \label{fig:initial_design}
\end{figure*}

Fig.~\ref{fig:initial_design} shows the initial design containing four circular holes within the entire meta-atom cell (only one hole in our quarter optimization domain), where the ratio $J_2/J_1 = 1$. Starting from this initial design, the final optimized designs for different values of $\omega$ are shown in the same figure. As $\omega$ increases, more emphasis is placed on minimizing $J_1$, which leads to more solid material being placed along the vertical direction. Conversely, as $\omega$ decreases, the designs that minimize $-J_2$ (or maximize $J_2$) have weak bars in the horizontal direction---thus maximizing energy release rates. Maximum anisotropy is attained for $\omega = 0.5$, \textit{i.e.}, $J_2/J_1 = 14.85$.

\appendix

\section{Appendix}
\subsection{Formulation}\label{sec:Q_matrix}

Following the work of Silva \textit{et al.}~\cite{Silva:2011} on  topological derivatives, the matrix $\boldsymbol Q$ in Eq.~\ref{eq:energy_release_rate} is given by
	\begin{align} 
\boldsymbol Q(\gamma, \beta) = \boldsymbol O(\beta)^{\intercal} \boldsymbol P(\gamma)^{\intercal} \boldsymbol {H}^{\intercal} (\gamma)\boldsymbol{H}(\gamma) \boldsymbol P(\gamma) \boldsymbol O(\beta).
	\end{align}
In this equation, $\boldsymbol O(\beta)$ is a transformation matrix for translating global to local stresses along the boundary, defined as
	\begin{align}
 \boldsymbol O(\beta) =
  \begin{bmatrix}
   c^2 & sc & sc & s^2 \\
  -sc &  c^2 & -s^2   & sc \\
  -sc & -s^2 &  c^2  & sc \\
   s^2 & -sc & -sc & c^2
   \end{bmatrix},
\label{eq:localglobal}
	\end{align}
where $c = \cos\beta$ and $s = \sin\beta$.
$\boldsymbol P(\gamma)$ is used to obtain stresses in polar coordinates, and it is given by
\begin{align} 
\boldsymbol {P}(\gamma) =
  \begin{bmatrix}
    s^2 \gamma & \displaystyle -\frac{\sin 2}{2} & \displaystyle -\frac{\sin 2\gamma}{2}   & c^2 \\
    - s c & \displaystyle \frac{c^2 -s^2}{2}  & \displaystyle \frac{c^2 - s^2}{2}& s c \end{bmatrix},
\end{align}
where this time where $c = \cos\gamma$ and $s = \sin\gamma$.
As stated in the work of Beghini \textit{et al.}~\cite{Beghini:1999}, $\boldsymbol H (\gamma)$ is a $2 \times 2$ matrix associated with angle $\gamma$, with components
\begin{align} 
\begin{split}
H_{11}(\gamma)  =\sum_{i=1}^{6} & \left\{ \left[1-\tan (\gamma)^{2}\right] \cdot c_{i}^{(\mathrm{I}, \mathrm{1})} \cos [(i-1) \gamma] \right. \\
  &  \left.  -\frac{\sin (\gamma)}{\cos (\gamma)^{3}} \cdot c_{i}^{(\mathrm{I}, 2)} \sin (i \gamma)\right\} 
\end{split}
\end{align}
\begin{align} 
\begin{split}
H_{12}(\gamma)  =\sum_{i=1}^{6} & \left\{2 \tan (\gamma) \cdot c_{i}^{(\mathrm{I}, \mathrm{1})} \cos [(i-1) \gamma]  \right. \\
 & + \left. \frac{1}{\cos (\gamma)^{2}} \cdot c_{i}^{(\mathrm{I}, 2)} \sin (i \gamma)\right\}
\end{split}
\end{align}
\begin{align} 
\begin{split}
H_{21}(\gamma)  =\sum_{i=1}^{6} & \left\{\left[1-\tan (\gamma)^{2}\right] \cdot c_{i}^{(\mathrm{II}, 1)} \sin (i \gamma) \right. \\
    &  \left. - \frac{\tan (\gamma)}{\cos (\gamma)^{2}} \cdot c_{i}^{(\mathrm{II}, 2)} \cos [(i-1) \gamma]\right\}
\end{split}
\end{align}
\begin{align}
\begin{split}
H_{22}(\gamma) = \sum_{i=1}^{6} & \left\{2 \tan (\gamma) \cdot c_{i}^{(\mathrm{II}, 1)} \sin (i \gamma) \right. \\
    & \left.  + \frac{1}{\cos (\gamma)^{2}} \cdot c_{i}^{(\mathrm{II}, 2)} \cos [(i-1) \gamma]\right\},
\end{split}
\end{align}
where $c_{i}^{(\mathrm{I}, \mathrm{1})}$, $c_{i}^{(\mathrm{II}, \mathrm{1})}$, $c_{i}^{(\mathrm{I}, \mathrm{2})}$, and $c_{i}^{(\mathrm{II}, \mathrm{2})}$ are given in Tab.~\ref{table:c_factors}.
\begin{table}[!ht]
    \centering
    \begin{tabular}{c c c c c} \toprule
         i    &   $c_{i}^{(\mathrm{I}, 1)}$    &  $c_{i}^{(\mathrm{II}, 1)}$    & $c_{i}^{(\mathrm{I}, 2)}$  &  $c_{i}^{(\mathrm{II}, 2)}$ \\  \midrule
        1   &   -0.174856 & -0.198196 & -0.419098 & 0.478653	  \\
        2   &    1.393783 & 0.681479 & -0.197271 & -0.130868 \\
        3   &     -0.278259 & -0.282608 & -0.445897 & 0.663435\\
        4   &    0.240695 & 0.136522 & -0.050066 & -0.066599 \\
        5   &     -0.071883 & -0.041562 & -0.022856 & 0.183693 \\
        6   &   0.011246 & 0.006177 & 0.003281 & -0.006140\\ \bottomrule
    \end{tabular}.
    \caption{Data of parameters $c_{i}^{(\mathrm{I}, \mathrm{1})}$, $c_{i}^{(\mathrm{II}, \mathrm{1})}$, $c_{i}^{(\mathrm{I}, \mathrm{2})}$, and $c_{i}^{(\mathrm{II}, \mathrm{2})}$ in matrix $\boldsymbol H(\gamma)$.}
    \label{table:c_factors}
\end{table}

\subsection{Sensitivity}
\label{sec:Sensitivity}

The sensitivity of the objective function $J$ with respect to a design variable $\boldsymbol s$ is derived by using the adjoint variable method. The Lagrangian function of the objective constructed by using the adjoint vectors $\boldsymbol{\lambda}_1$ and $\boldsymbol{\lambda}_2$ is expressed as
	\begin{align}
	L = J + \boldsymbol \lambda_1^{\intercal} (\boldsymbol K_1  \boldsymbol {U}_1 -\boldsymbol{F}_1) + \boldsymbol \lambda_2^{\intercal} (\boldsymbol K_2  \boldsymbol {U}_2 -\boldsymbol{F}_2).
	\end{align} 
Then the derivative of $L$ with respect to the $j$th design variable ${s}_j$ is given by
\begin{align} 
\begin{split}
\frac{\mathrm{d} L}{\mathrm{d} s_{j}} 
& =\omega \frac{\partial J_1}{\partial s_{j}} - (1-\omega) \frac{\partial J_2}{\partial s_{j}}\\
&+\boldsymbol{\lambda}_1^{\intercal}\frac{\partial \boldsymbol K_1}{\partial s_{j}} \boldsymbol{U}_1-\boldsymbol{\lambda}_1^{\intercal}\frac{\partial \boldsymbol{F}_1}{\partial s_{j}} \\
& +\boldsymbol{\lambda}_2^{\intercal}\frac{\partial \boldsymbol K_2}{\partial s_{j}} \boldsymbol{U}_2-\boldsymbol{\lambda}_2^{\intercal}\frac{\partial \boldsymbol{F}_2}{\partial s_{j}}\\
& +\left(\omega \frac{\partial J_1}{\partial \boldsymbol{ U}_1} +\boldsymbol{\lambda}_1^{\intercal}\boldsymbol K_1 \right)\frac{\partial \boldsymbol{U}_1}{\partial s_{j}} \\
& + \left(-(1-\omega) \frac{\partial J_2}{\partial \boldsymbol{U}_2}+\boldsymbol{\lambda}_2^{\intercal}\boldsymbol K_2 \right)\frac{\partial \boldsymbol{U}_2}{\partial s_{j}}.
\end{split}
\end{align}
After obtaining the adjoint vectors $\boldsymbol{\lambda}_1$ and $\boldsymbol{\lambda}_2$ by solving the following adjoint equations
\begin{align} 
\omega \frac{\partial J_1}{\partial \boldsymbol{ U}_1} +\boldsymbol{\lambda}_1^{\intercal}\boldsymbol K_1 = \boldsymbol 0,
\end{align}
and
\begin{align}
-(1-\omega) \frac{\partial J_2}{\partial \boldsymbol{U}_2}+\boldsymbol{\lambda}_2^{\intercal}\boldsymbol K_2 = \boldsymbol 0,
\end{align}
the above sensitivity formulation is simplified as
\begin{align} 
\begin{split}
\frac{\mathrm{d} L}{\mathrm{d} s_{j}} 
& =\omega \frac{\partial J_1}{\partial s_{j}} - (1-\omega) \frac{\partial J_2}{\partial s_{j}}\\
&+\boldsymbol{\lambda}_1^{\intercal}\frac{\partial \boldsymbol K_1}{\partial s_{j}} \boldsymbol{U}_1-\boldsymbol{\lambda}_1^{\intercal}\frac{\partial \boldsymbol{F}_1}{\partial s_{j}} \\
& +\boldsymbol{\lambda}_2^{\intercal}\frac{\partial \boldsymbol K_2}{\partial s_{j}} \boldsymbol{U}_2-\boldsymbol{\lambda}_2^{\intercal}\frac{\partial \boldsymbol{F}_2}{\partial s_{j}}.
\end{split}
\end{align}

\newpage

\end{document}